\documentclass[12pt,a4paper,final]{iopart}

\usepackage{iopams}  
\usepackage{graphicx}
\usepackage[breaklinks=true,colorlinks=true,linkcolor=blue,urlcolor=blue,citecolor=blue]{hyperref}

\usepackage{subcaption}
\begin{document}
\title[The prisoner's dilemma as a cancer model]{The prisoner's dilemma as a cancer model}

\author{Jeffrey West}
\address{Dept. of Aerospace \& Mechanical Engineering, University of Southern California, Los Angeles, CA}
\ead{westjb@usc.edu}

\author{Zaki Hasnain}
\address{Dept. of Aerospace \& Mechanical Engineering, University of Southern California, Los Angeles, CA}
\ead{zhasnain@usc.edu}

\author{Jeremy Mason}
\address{Dept. of Biological Sciences, University of Southern California, Los Angeles, CA}
\ead{masonj@usc.edu}

\author{Paul K. Newton}
\address{Dept. of Aerospace \& Mechanical Engineering, Dept. of Mathematics, Norris Comprehensive Cancer Center, University of Southern California, Los Angeles, CA}
\ead{newton@usc.edu}

\begin{abstract}
Tumor development is an evolutionary process in which a heterogeneous population of cells with differential growth capabilities compete for resources in order to gain a proliferative advantage. What are the minimal ingredients needed to recreate some of the emergent features of such a developing complex ecosystem? What is a tumor doing before we can detect it? We outline a mathematical model, driven by a stochastic Moran process, in which cancer cells and healthy cells compete for dominance in the population. Each are assigned payoffs according to a Prisoner's Dilemma evolutionary game where the healthy cells are the cooperators and the cancer cells are the defectors. With point mutational dynamics, heredity, and a fitness landscape controlling birth and death rates, natural selection acts on the cell population and simulated `cancer-like' features emerge, such as Gompertzian tumor growth driven by heterogeneity, the log-kill law which (linearly) relates therapeutic dose density to the (log) probability of cancer cell survival, and the Norton-Simon hypothesis which (linearly) relates tumor regression rates to tumor growth rates. We highlight the utility, clarity, and power that such models provide, despite (and because of) their simplicity and built-in assumptions.
\end{abstract}


\section{Introduction}
Cancer is an evolutionary process taking place within a genetically and functionally heterogeneous population of cells that traffic from one anatomical site to another via hematogenous and lymphatic routes \cite{bib1,bib7,bib12,bib53,bib60}. The population of cells associated with the primary and metastatic tumors evolve, adapt, proliferate, and disseminate in an environment in which a fitness landscape controls survival and replication \cite{bib31}. Tumorigenesis occurs as the result of inherited and acquired genetic, epigenetic and other abnormalities accumulated over a long period of time in otherwise normal cells \cite{bib28,bib49}.  Before we can typically detect the presence of a tumor, the cells are already competing for resources in a Darwinian struggle for existence in tissues that progressively age and evolve. It is well established that the regenerative capacity of individual cells within a tumor, and their ability to traffic multi-directionally from the primary tumor to metastatic tumors all represent significant challenges associated with the efficacy of different cancer treatments and our resulting ability to control systemic spread of many soft-tissue cancers \cite{bib36,bib59}. Details of the metastatic and evolutionary process are poorly understood, particularly in the subclinical stages when tumors are actively developing but not yet clinically visible \cite{bib52}. It could be argued that in order to truly understand cancer progression at the level in which quantitative predictions become feasible, it is necessary to understand how genetically and epigenetically heterogeneous populations of cells compete and evolve within the tumor environment well before the tumor is clinically detectable. Additionally, a better understanding of how these populations develop resistance to specific therapies \cite{bib16,bib22} might help in developing optimal strategies to attack the tumor and slow disease progression.

Evolutionary game theory is perhaps the best quantitative framework for modeling evolution and natural selection. It is a dynamic version of classical game theory in which a game between two (or more) competitors is played repeatedly, giving each participant the ability to adjust their strategy based on the outcome of the previous string of games. While this may seem like a minor variant of classical (static) game theory, as developed by the mathematicians von Neumann and Morgenstern in the 1940's \cite{bib57}, it is not. Developed mostly by the mathematical biologists John Maynard Smith and George Price in the 1970s \cite{bib29,bib30} and Martin Nowak and Karl Sigmund \cite{bib44,bib47} more recently, this dynamic generalization of classical game theory has proven to be one of the main quantitative tools available to evolutionary biologists (with a mathematical bent) whose goal is to understand natural selection in evolving populations. In this biological context, a strategy is not necessarily a deliberate course of action, but an inheritable trait \cite{bib50}. Instead of identifying Nash equilibria, as in the static setting \cite{bib34,bib35}, one looks for evolutionary stable strategies (ESS) and fixation probabilities \cite{bib19,bib44} of a subpopulation. This subpopulation might be traced to a specific cell with enhanced replicative capacity, for example, that has undergone a sequence of mutations and is in the process of clonally expanding \cite{bib48}. A relevant question in that case is what is the probability of fixation of that subpopulation? More explicitly, how does one subpopulation invade another in a developing colony of cells? 

One game in particular, the Prisoner's Dilemma game, has played a central role in cancer modeling (as well as other contexts such as political science and economics) \cite{bib2,bib3,bib4,bib10,bib11,bib14,bib15,bib17,bib18,bib19,bib20,bib21,bib23,bib45,bib46,bib47,bib54,bib55,bib56,bib58}. It was originally developed by Flood, Dresher and Tucker in the 1950s as an example of a game which shows how rational players might not cooperate, even if it seems to be in their best interest to do so. The evolutionary version of the Prisoner's Dilemma game has thus become a paradigm for the evolution of cooperation among a group of selfish individuals and thus plays a key role in understanding and modeling the evolution of altruistic behavior \cite{bib2,bib3}. Perhaps the best introductory discussion of these ideas is found in Dawkins' celebrated book, The Selfish Gene \cite{bib8}. The framework of evolutionary game theory allows the modeler to track the relative frequencies of competing subpopulations with different traits within a bigger population by defining mutual payoffs among pairs within the group. One can then define a fitness landscape over which the subpopulations evolve. The fitness of different phenotypes is frequency dependent and is associated with reproductive prowess, while the `players' in the evolutionary game compete selfishly for the largest share of descendants \cite{bib19,bib58}.  Our goal in this article is provide a brief introduction to how the Prisoner's Dilemma game can be used to model the interaction of competing subpopulations of cells, say healthy, and cancerous, in a developing tumor and beyond.

\section{The prisoner's dilemma evolutionary game}
An evolutionary game between two players is defined by a 2 x 2 payoff matrix which assigns a reward to each player (monetary reward, vacation time, reduced time in jail, etc.) on a given interaction. Let us call the two players A and B. In the case of a prisoner's dilemma game between cell types in an evolving population of cells, let there be two subpopulations of cell types which we will call `healthy', and `cancerous'. We can think of the healthy cells as the subpopulation that is cooperating, and the cancer cells as formerly cooperating cells that have defected via a sequence of somatic driver mutations.  Imagine a sequence of `games' played between two cells (A and B) selected at random from the population, but chosen in proportion to their prevalence in the population pool. Think of a cancer-free organ or tissue as one in which a population of healthy cells are all cooperating, and the normal organ functions are able to proceed, with birth and death rates that statistically balance, so an equilibrium healthy population is maintained (on average). Now imagine a mutated cell introduced into the population with enhanced proliferative capability as encoded by its genome as represented as a binary sequence of 0's and 1's carrying forward its genetic information (which is passed on to daughter cells). A schematic diagram associated with this process is shown in Figure 1. We can think of this cancer cell as a formerly cooperating cell that has defected and begins to compete against the surrounding population of healthy cells for resources and reproductive prowess. From that point forward, one can imagine tumor development to be a competition between two distinct competing subpopulations of cells, healthy (cooperators) and cancerous (defectors). We are interested in the growth rates of a `tumor' made up of a collection of cancer cells within the entire population, or equivalently, we are interested in tracking the proportion of cancer cells, $i(t)$, vs. the proportion of healthy cells, $N-i(t)$, in a population of $N$ cells comprising the simulated tissue region. 

To quantify how the interactions proceed, and how birth/death rates are ultimately assigned, we introduce the 2 x 2 prisoner's dilemma payoff matrix:

\begin{equation} \label{eqn1}
A = \left( \begin{array}{cc}
a & b \\
c & d \end{array} \right) = \left( \begin{array}{cc}
3 & 0 \\
5 & 1 \end{array} \right) 
\end{equation}

The essence of the prisoner's dilemma game is the two players compete against each other, and each has to decide what best strategy to adopt in order to maximize their payoff. This 2 x 2 matrix assigns the payoff (e.g. reward) to each player on each interaction. My options, as a strategy or, equivalently, as a cell type, are listed along the rows, with row 1 associated with my possible choice to cooperate, or equivalently my cell type being healthy, and row 2 associated with my possible choice to defect, or equivalently my cell type being cancerous. Your options are listed down the columns, with column 1 associated with your choice to cooperate (or you being a healthy cell), and column 2 associated with your choice to defect (or you being a cancer cell). The analysis of a rational player in a prisoner's dilemma game runs as follows. I do not know what strategy you will choose, but suppose you choose to cooperate (column 1). In that case, I am better off defecting (row 2) since I receive a payoff of 5 instead of 3 (if I also cooperate). Suppose instead you choose to defect (column 2). In that case, I am also better off defecting (row 2) since I receive a payoff of 1 instead of 0 (if I were to have cooperated). Therefore, {\em no matter what you choose, I am better off (from a pure payoff point of view) if I defect}. What makes this game such a useful paradigm for strategic interactions ranging from economics, political science, biology, and even psychology \cite{bib2,bib29,bib58} is the following additional observation. {\em You will analyze the game in exactly the same way I did (just switch the roles of me and you in the previous rational analysis), so you will also decide to defect no matter what I do}. The upshot if we both defect is that we will each receive a payoff of 1, instead of each receiving a payoff of 3 if we had both chosen to cooperate. The defect-defect combination is a Nash equilibrium \cite{bib34,bib35}, and yet it is sub-optimal for both players and for the system as a whole. Rational thought rules out the cooperate-cooperate combination which would be better for each player (3 points each) and for both players combined (6 points). In fact, the Nash equilibrium strategy of defect-defect is the worst possible system wide choice, yielding a total payoff of 2 points, compared to the cooperate-defect or defect-cooperate combination, which yields a total payoff of 5 points, or the best system-wide strategy of cooperate-cooperate yielding a total payoff of 6 points.

The game becomes even more interesting if it is played repeatedly \cite{bib58}, stochastically \cite{bib55}, and with spatial structure \cite{bib27} with each player allowed to decide what strategy to use on each interaction so as to accumulate a higher payoff than the competition over a sequence of $N$ games. In order to analyze this kind of an evolving set-up, a fitness function must be introduced based on the payoff matrix A. Let us now switch our terminology so that the relevance to tumor cell kinetics becomes clear. When modeling cell competition, one has to be careful about the meaning of the term `choosing a strategy'. Cells do not choose a strategy, but they do behave in different ways depending on whether they are normal healthy cells cooperating as a cohesive group, with birth and death rates that statistically balance, or whether they are cancer cells with an overactive cell division mechanism (as triggered by the presence of oncogenes) and an underactive `break' mechanism (as triggered by the absence of tumor suppressor genes) \cite{bib60}. In our context, it is not the strategies that evolve, as cells cannot change type based on strategy (only based on mutations), but the prevalence of each cell type in the population is evolving, with the winner identified as the sub-type that first saturates in the population.

\section{A tumor growth model}

\begin{figure}[ht!]
\begin{subfigure}{.9\textwidth}
\begin{center}
  \noindent \includegraphics[width=0.4\linewidth]{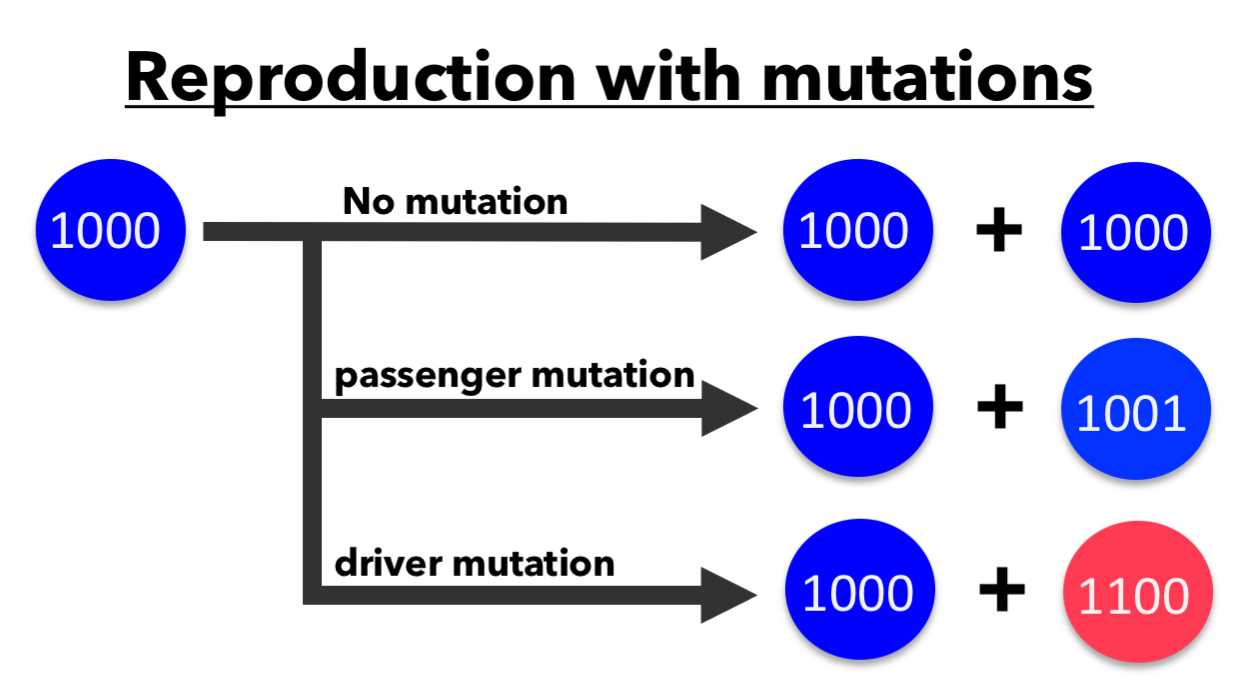}
  \caption{}{}
  \label{fig1:a}
\end{center}
\end{subfigure}
\newline
\begin{subfigure}{0.9\textwidth}
\begin{center}
  \noindent \includegraphics[width=0.7\linewidth]{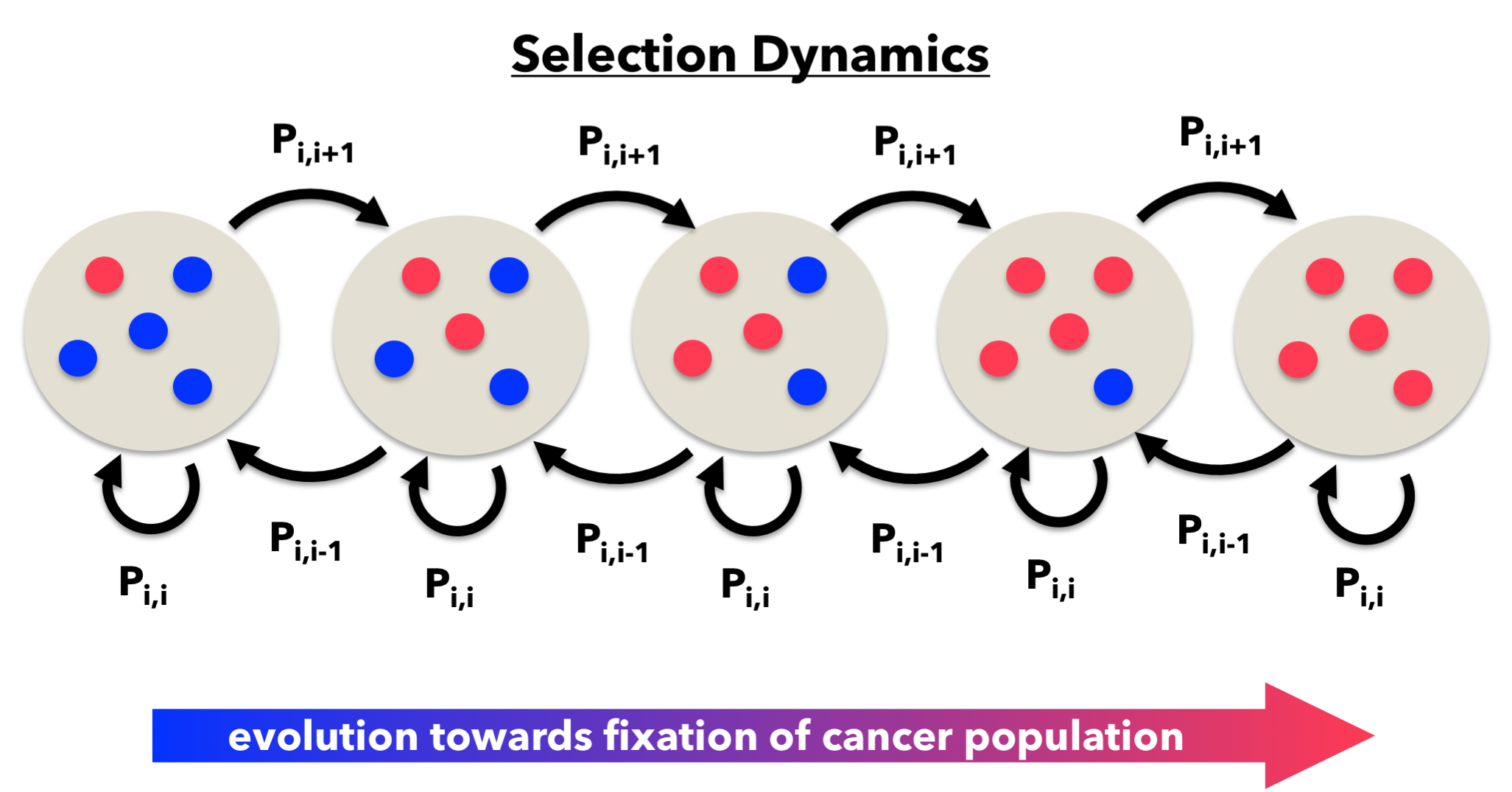}
  \caption{}{}
  \label{fig1:b}
\end{center}
\end{subfigure}
\caption{Schematic of the Moran Process --- (a) During each time step, a single cell is chosen for reproduction, where an exact replica is produced. With probability $m$ ($0 \leq m \leq 1$), a mutation may occur. (b) The number of cancer cells, $i$, is defined on the state space $i = 0, 1, \ldots, N$ where $N$ is the total number of cells. The cancer population can change at most by one each time step, so a transition exists only between state $i$ and $i – 1$, $i$, and $i + 1$.}
\label{fig1:fig1}
\end{figure}

Consider a population of $N$ cells driven by a stochastic birth-death process as depicted in Figure \ref{fig1:fig1}, with red cells depicting cancer cells (higher fitness) and blue cells depicting healthy cells (lower fitness, but cooperative). We model the cell population as a stochastic Moran process \cite{bib61} of $N$ cells, `$i$' of which are cancerous, `$N-i$' of which are healthy. If each cell had equal fitness, the birth-death rates would all be equal and a statistical balance would ensue. At each step, a cell is chosen (randomly but based on the prevalence in the population pool) and eliminated (death), while another is chosen to divide (birth). If all cells had equal fitness, the birth/death rates of the cancer cells would be $i/N$, while those of the healthy cells would be $(N-i)/N$.  With no mechanism for introducing a cancer cells in the population, the birth/death rates of the healthy cells would be 1, and no tumor would form. 

\begin{figure}[!ht]
\begin{center}
\includegraphics[width=0.7\textwidth]{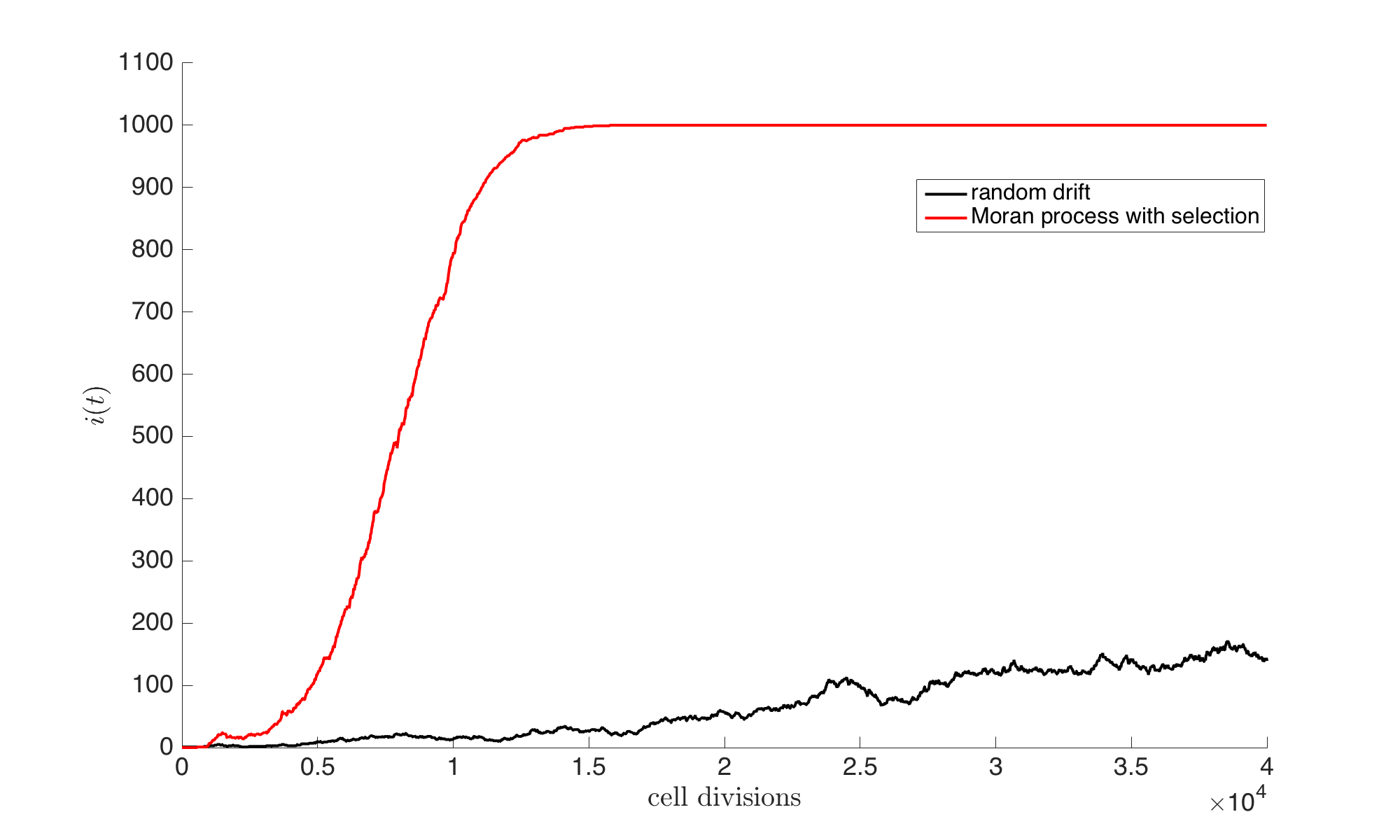}
\end{center}
\caption{Emergence of Gompertzian growth via selection --- Random drift (black) plotted for a single simulation of $10^3$ cells for $4\cdot 10^4$ generations shows no particular shape. A single simulation of the Moran process (red) with selection ($w = 0.5$) and mutations ($m = 0.1$) gives rise to the characteristic S-shaped curve associated with Gompertzian growth.}
\label{fig2}
\end{figure}

Now, introduce one cancer cell into the population of healthy cells, as shown in Figure \ref{fig1:fig1}. At each step, there would be a certain probability of this cell dividing ($P_{i,i+1}$), being eliminated ($P_{i,i-1}$), or simply not being chosen for either division or death ($P_{i,i}$). Based on this random process, it might be possible for the cancer cells to saturate the population, as shown by one simulation in Figure \ref{fig2} depicting $N=1000$ cells, with initially $i=1$ cancer cell, and $N-i = 999$ healthy cells. However, the growth curve would not show any distinct shape (Figure \ref{fig2} (black)), and might well become extinct after any number of cell divisions, as opposed to reaching saturation. But we emphasize that without mutational dynamics, heritability, and natural selection operating on the cell population, the shape of the growth curve would look random, and we know this is not how tumors tend to grow \cite{bib25,bib26}. By contrast, Figure \ref{fig2} (red) shows a Gompertzian growth curve starting with exponential growth of the cancer cell subpopulation, followed by linear growth, ending with saturation. The growth rate is not constant throughout the full history of tumor development, but after an initial period of exponential growth, the rate decelerates until the region saturates with cancer cells. The basic ingredients necessary to sustain Gompertzian growth seem to be: an underlying stochastic engine of developing cells, mutational dynamics, heritability, and a fitness landscape that governs birth and death rates giving rise to some sort of natural selection.

\subsection{Mutations and heritability}

\begin{figure}[!ht]
\begin{center}
\includegraphics[width=0.5\textwidth]{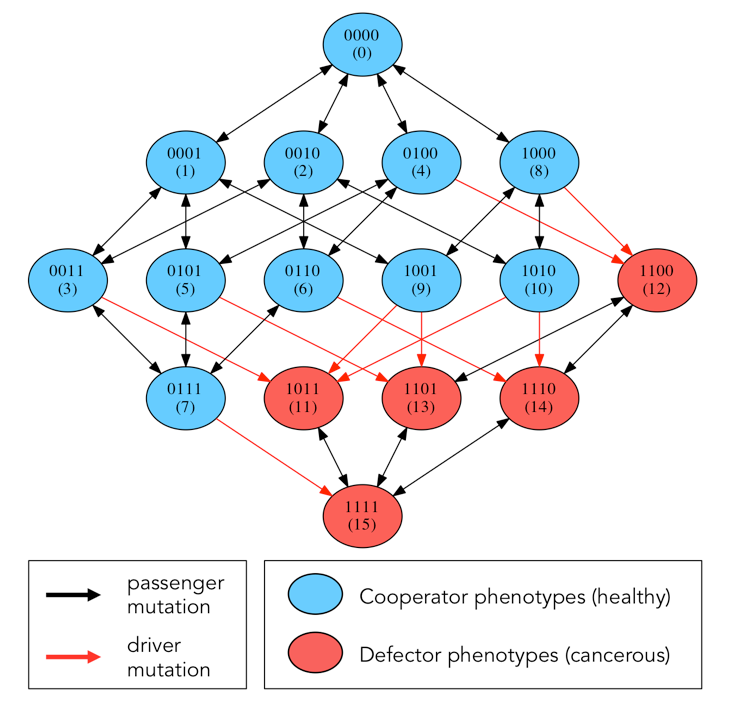}
\end{center}
\caption{Markov Point Mutation Diagram --- Diagram shows 16 genetic cell types based on 4-digit binary string and the effect of a point mutation on each cell type. Blue indicates healthy cell type (0000 --- 1010), red indicates cancerous cell type (1011 --- 1111). Black arrows indicate passenger mutations (healthy to healthy or cancer to cancer), red arrows indicate driver mutations (healthy to cancer).}
\label{fig3}
\end{figure}

Each of the $N$ cells in our simulated population carries with it a discrete packet of information that represents some form of molecular differences among the cells. In our model, we code this information in the form of a 4-digit binary string from 0000 up to 1111, giving rise to a population made up of 16 distinct cell types. At each discrete step in the birth-death process, one of the digits in the binary string is able to undergo a point mutation \cite{bib13,bib28}, where a digit spontaneously flips from 0 to 1, or 1 to 0, with probability $p_m$. The mutation process is shown in Figure \ref{fig1:fig1}, while a mutation diagram is shown in Figure \ref{fig3} in the form of a directed graph. This figure shows the possible mutational transitions that can occur in each cell, from step to step in a simulation. A typical simulation begins with a population of $N$ healthy cells, all with identical binary strings 0000. The edges on the directed graph represent possible mutations that could occur on a given step. The first 11 binary string values (0-10) represent healthy cells in our model that are at different stages in their evolutionary progression towards becoming a cancer cell (the exact details of this genotype to phenotype map do not matter much). Mutations strictly within this subpopulation are called passenger mutations as the cells all have the same fitness characteristics. The first driver mutation occurs when a binary string reaches value 11-15. The first cell that transitions from the healthy state to the cancerous state is the renegade cell in the population that then has the potential to clonally expand and take over the population. How does this process occur?

\subsection{The fitness landscape}

At the heart of how the Prisoner's Dilemma evolutionary game dictates birth and death rates which in turn control tumor growth, is the definition of cell fitness. Let us start by laying out the various probabilities of pairs of cells interacting and clearly defining payoffs when there are i cancer cells, and $N-i$ healthy cells in the population. The probability that a healthy cell interacts with another healthy cell is given by $(N-i-1)/(N-1)$, whereas the probability that a healthy cell interacts with a cancer cell is $i/(N-1)$. The probability that a cancer cell interacts with a healthy cell is $(N-i)/(N-1)$, whereas the probability that a cancer cell interacts with another cancer cell is $(i-1)/(N-1)$. 

In a fixed population of $N$ cells, with i cancer cells, the number of healthy cells is given by $N-i$. The average payoff of a single cell ($\pi^H, \pi^C$), is dependent on the payoff matrix value weighted by the relative frequency of types in the current population:

\begin{equation}\label{eqn2}
\pi_{i}^H = \frac{a(N-i-1) + bi}{N-1}
\end{equation}

\begin{equation}\label{eqn3}
\pi_{i}^C = \frac{c(N-i) + d(i-1)}{N-1}
\end{equation}

\begin{figure}[ht!]
\begin{subfigure}{.9\textwidth}
  \begin{center}
  \noindent \includegraphics[width=0.8\linewidth]{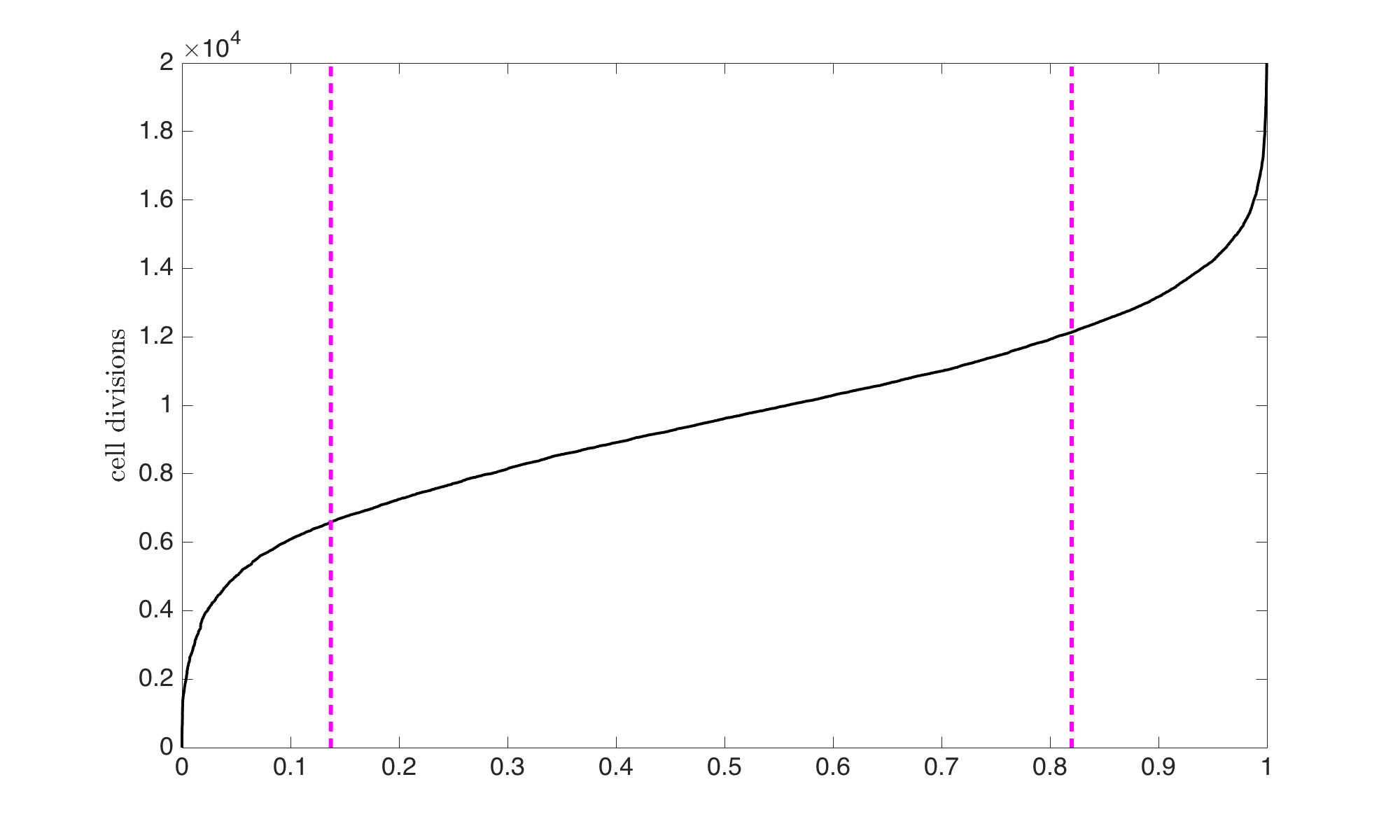}
  \end{center}
  \caption{}{}
  \label{fig4:a}
\end{subfigure}
\newline
\begin{subfigure}{.9\textwidth}
  \begin{center}
  \noindent \includegraphics[width=0.8\linewidth]{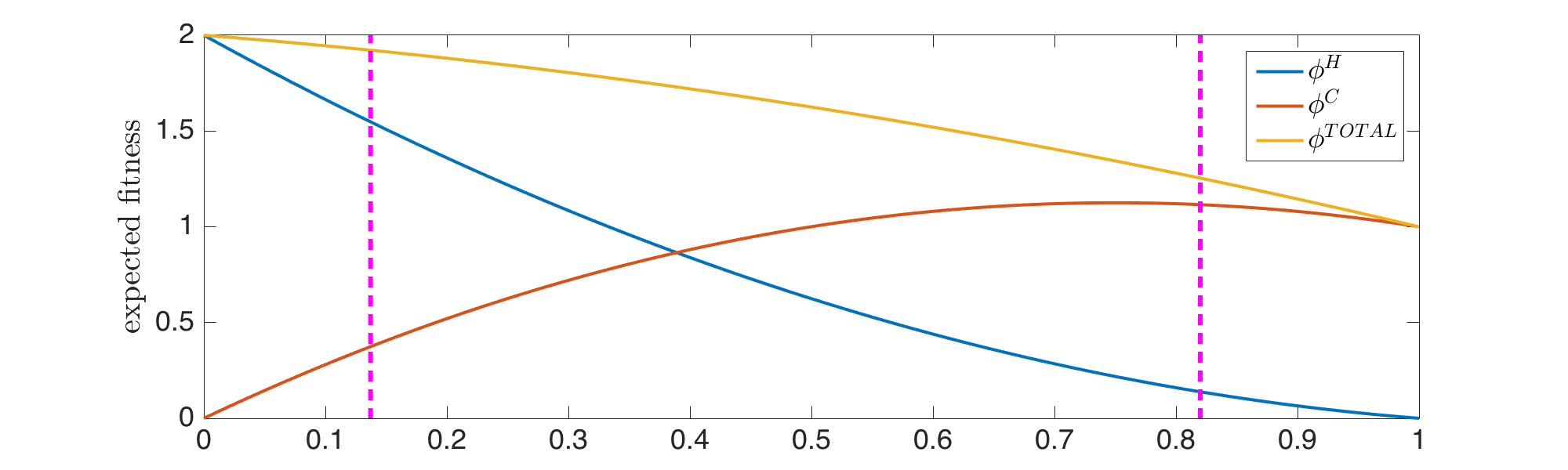}
  \end{center}
  \caption{}{}
  \label{fig4:b}
\end{subfigure}
\newline
\begin{subfigure}{.9\textwidth}
  \begin{center}
  \noindent \includegraphics[width=0.8\linewidth]{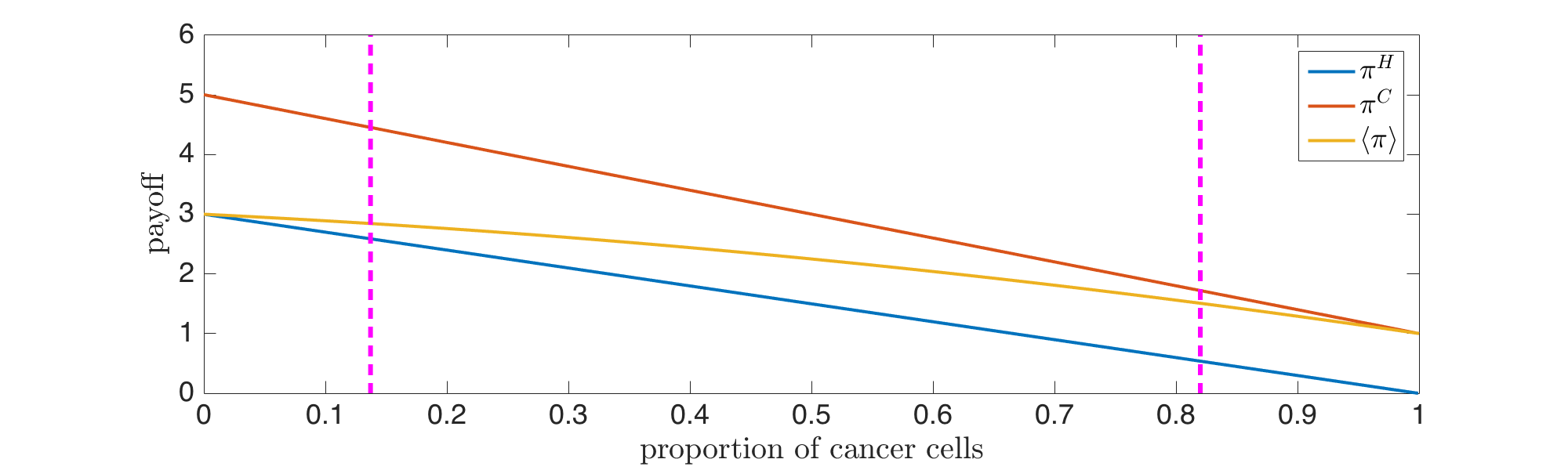}
  \end{center}
  \caption{}{}
  \label{fig4:c}
\end{subfigure}
\caption{Tumor fitness drives tumor growth --- (a) The average of 25 stochastic simulations ($N = 1000$ cells, $w= 0.5$, $m = 0.1$) is plotted for 20,000 cell divisions to show the cancer cell population (defectors) saturating. The pink lines delineate the regions of tumor growth (defined by the maximum and minimum points of the second-derivative of $i(t)$). (b) Fitness of the healthy population, cancer population, and total population plotted for the range cancer cell proportion. (c) Average payoff of a single healthy cell, cancer cell, and all cells plotted for the range cancer cell proportion.}
\label{fig4:fig4}
\end{figure}

\noindent Here, $a=3$, $b=0$, $c=5$, $d=1$ are the entries in the Prisoner's Dilemma payoff matrix (\ref{eqn1}). For the Prisoner's dilemma game, the average payoff of a single cancer cell is always greater than the average payoff for a healthy cell (Figure \ref{fig4:c}). With the invasion of the first cancer cell, the higher payoff gives a higher probability of survival when in competition with a single healthy cell. 

Selection acts on the entire population of cells as it depends not on the payoff, but on the effective fitness of the subtype population. The effective fitness of each cell type ($f^H$, $f^C$) is given by the relative contribution of the payoff of that cell type, weighted by the selection pressure: 

\begin{equation}\label{eqn4}
f_{i}^H = 1-w +w \pi_{i}^H
\end{equation}

\noindent and the fitness of the cancer cells as:

\begin{equation}\label{eqn5}
f_{i}^C = 1-w +w \pi_{i}^C
\end{equation}
   
\noindent The probability of birthing a new cancer cell depends on the relative frequency (random drift) weighted by the effective fitness, and the death rate is proportional to the relative frequency. The transition probabilities can be written:

\begin{equation}\label{eqn6}
P_{i,i+1} = \frac{if_i^C}{if_i^C + (N-i)f_i^H}\frac{N-i}{N}
\end{equation}

\begin{equation}\label{eqn7}
P_{i,i-1} = \frac{(N-i)f_i^H}{if_i^C + (N-i)f_i^H}\frac{i}{N}
\end{equation}

\begin{equation}\label{eqn8}
P_{i,i} = 1 - P_{i,i+1} - P_{i,i-1}; \quad P_{0,0} = 1; \quad P_{N,N} = 1.
\end{equation}

In the event of the first driver mutation, the first cancer cell is birthed. At the beginning of the simulation, the effective fitness of the healthy population is much greater than the fitness of the cancer population (Figure \ref{fig4:b}). This is because although the single cancer has a higher {\em payoff} than any of the healthy cells, the number of healthy cells far outnumber the single cancer cells. That single cancer cell initiates a regime of explosive high growth and the fitness of the cancer population steadily increases. Cancer cells are continually competing with healthy cells and receiving a higher payoff in this regime (compare the payoff entries of a cancer cell receiving $c = 5$ vs a healthy cell receiving $b = 0$). At later times, growth slows because cancer cells are competing in a population consisting mostly of other cancer cells. The payoff for a cancer cell is dramatically lower when interacting with a cancer cell (observe the payoff entry of both cancer cells receiving $d = 1$ when interacting). As the cancer population grows, the payoff attainable decreases and growth slows. In addition, the average fitness of the total population steadily declines because each interaction derives less total payoff, from $c + b = 5$ to $d + d = 1$.

\begin{figure}[!ht]
\begin{center}
\includegraphics[width=0.7\textwidth]{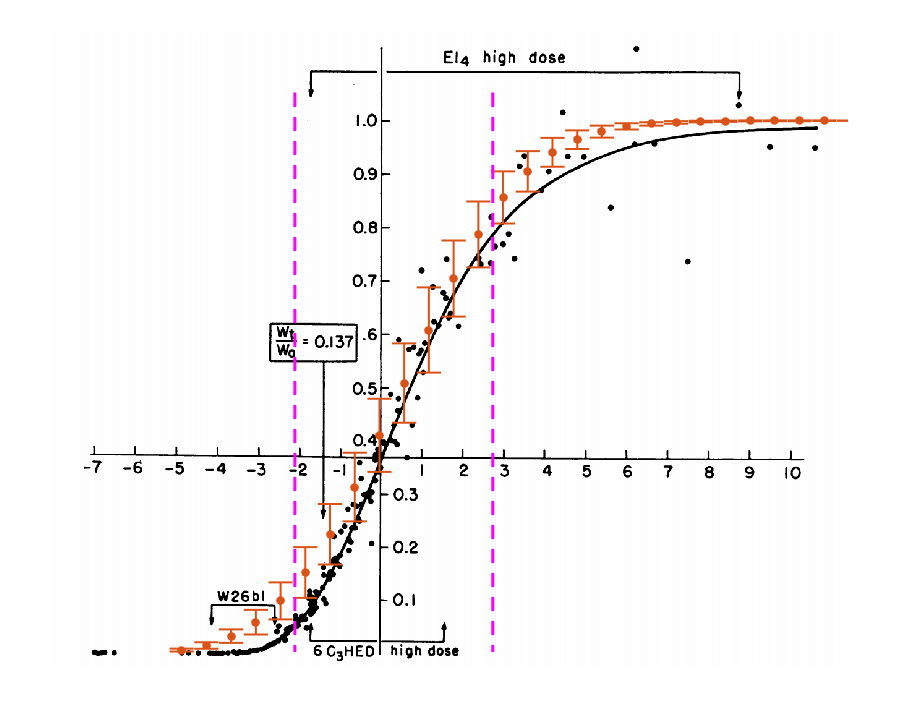}
\end{center}
\caption{Moran Process fit to Gompertzian Growth Data --- The mean and deviation of 25 stochastic simulations ($N = 10^3$ cells, $w= 0.7$, $m = 0.3$) is overlaid on data from a ``normalized" Gompertzian \cite{bib25,bib26}. Values for m and w were chosen by implementing a least-squares fit to the data over a range of $m$ ($0 \leq m \leq 1$),  and $w$ ($0 \leq w \leq 1$). Pink lines delineate regions of growth (defined by the maximum and minimum points of the second-derivative of $i(t)$).}
\label{fig5}
\end{figure}

This complex process of competition among cell types and survival of subpopulations, where defection is selected over cooperation, produces a Gompertzian growth curve shown in Figure \ref{fig5}, and compared with a compilation based on a wide range of data first shown in \cite{bib25,bib26}. It is now well established that tumor cell populations (and other competing populations, such as bacteria and viral populations) generally follow this growth pattern, although the literature is complicated by the fact that different parts of the growth curve have vastly different growth rates \cite{bib25,bib26}, and it is nearly impossible to follow the growth of a population of cancer cells {\em in vivo} from the first cancer cell through to an entire tumor made up of $O(10^9 - 10^{12})$ cells. Growth rates are typically measured for a short clinical time period \cite{bib25,bib26}, and then extrapolated back to the first renegade cell, and forward to the fully developed tumor population.

\subsection{Heterogeneity drives growth}

Insights into the process by which growth rates vary and conspire to produce a Gompertzian shape can be achieved by positing that growth is related to molecular and cellular heterogeneity of the developing population \cite{bib5,bib24,bib53}. Indeed, an outcome of the model is that molecular heterogeneity (i.e. the dynamical distribution of the 4-digit binary string 0000---1111 making up the population of cells) drives growth. Consider entropy \cite{bib6,bib39} of the cell population as a measure of heterogeneity:

\begin{equation}
E(t) = - \sum_{i=1}^{N} p_i \log_{2} p_i \label{eqn9}
\end{equation}

\noindent (here, log is defined as base 2). The probability $p_i$ measures the proportion of cells of type $i$, with $i = 1,\ldots,16$ representing the distribution of binary strings ranging from 0000 to 1111. We typically course-grain this distribution further so that cells having strings ranging from 0000 up to 1010 are called `healthy', while those ranging from 1011 to 1111 are `cancerous'. Then growth is determined by:

\begin{equation}
\frac{dn_{E}}{dt}= \alpha E(t) \label{eqn10}
\end{equation}

It follows from (\ref{eqn10}) that the cancer cell proportion $n_E(t)$ can be written in terms of entropy as:

\begin{equation}
n_{E}(t) = \alpha \int_{0}^t E(t) dt \label{eqn11}
\end{equation}

This relationship between growth of the cancer cell population and entropy is pinned down and detailed in \cite{bib61}. We consider it to be one of the key emergent features of our simple model. 

\begin{figure}[!ht]
\begin{center}
\includegraphics[width=0.7\textwidth]{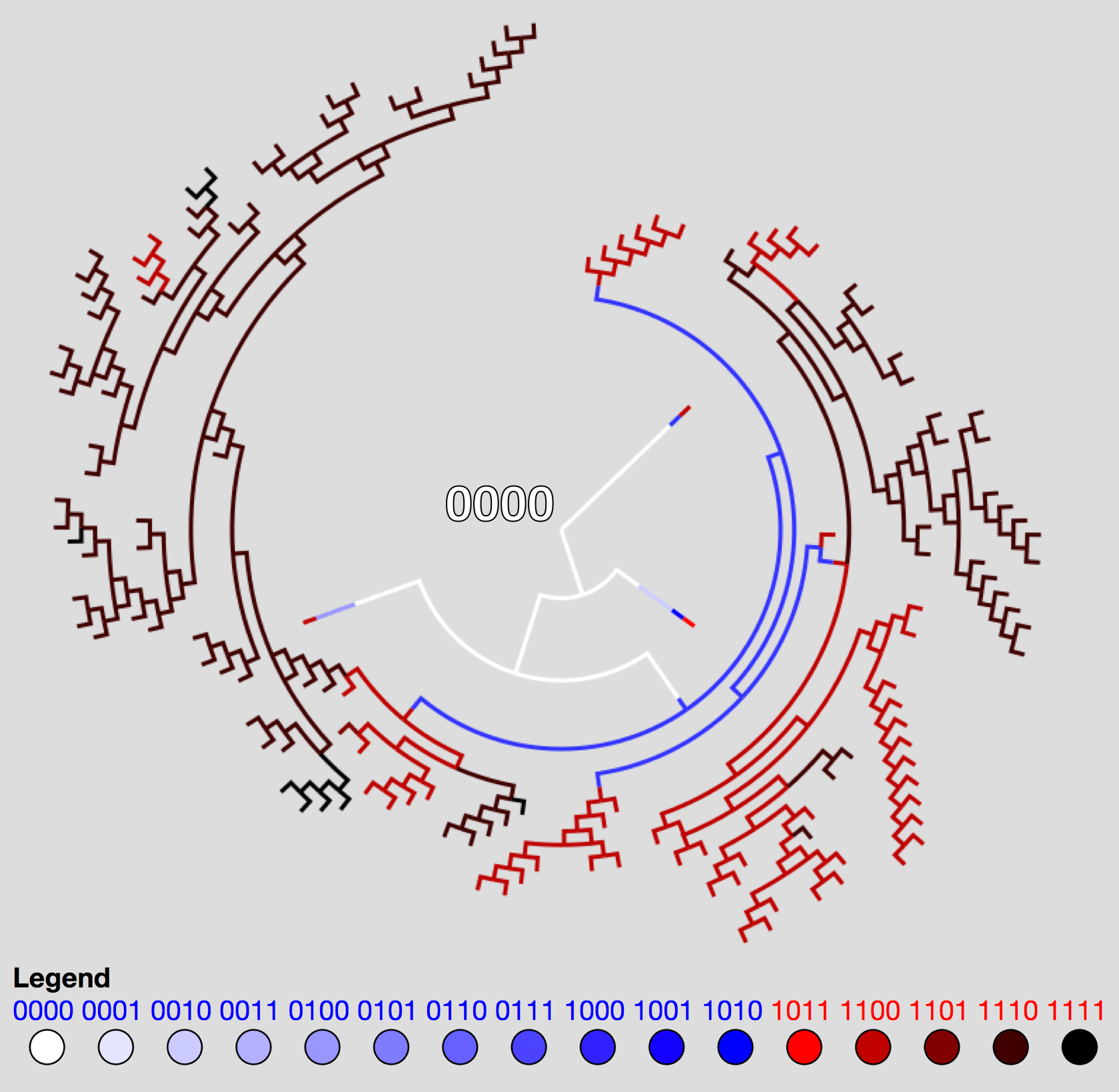}
\end{center}
\caption{Phylogenetic Tree --- Sample dendritic phylogenetic tree tracking point mutations as time extends radially, depicting the emergence of molecular heterogeneity. The tree shows a simulation of 30 cells (all with genetic string 0000 at the beginning of the simulation) with strong selection ($w = 1$, $m = 0.2$). Pathways are color coded to indicate genetic cell type.}
\label{fig6}
\end{figure}

A typical example of the emergence of genetic heterogeneity in our model system is shown in the form of a phylogenetic tree in Figure \ref{fig6}. This particular tree is obtained via a simulation of only 30 healthy phenotypic cells (0000), which during the course of a simulation expand out (radially in time) to form a much more heterogeneous population of cells at the end of the simulation. In our model, the genetic time-history of each cell is tracked and the population can be statistically analyzed after the simulation finishes.

\section{Simulated drug dosing strategies and therapeutic response}

\begin{figure}[!ht]
\begin{center}
\includegraphics[width=0.9\textwidth]{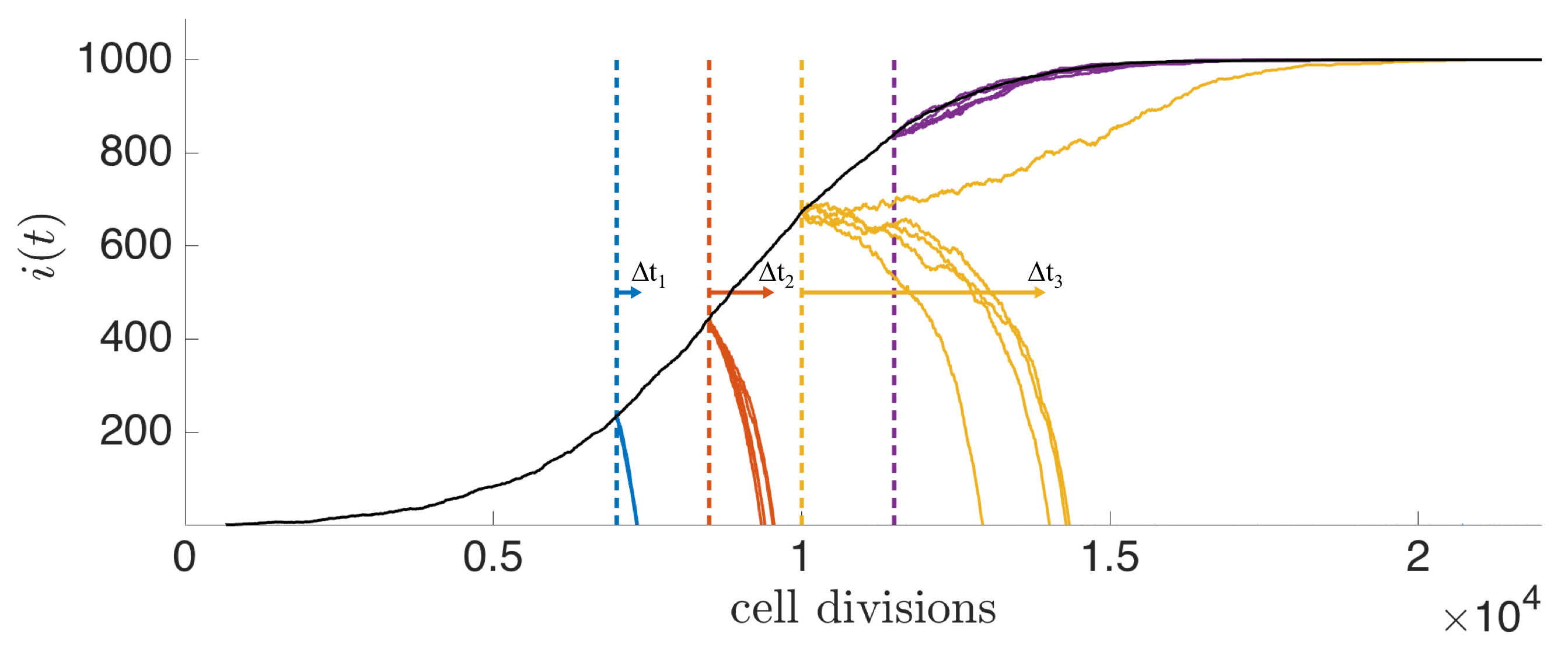}
\end{center}
\caption{Effects of early stage, mid-stage, and late-stage therapies --- An average of 5 stochastic simulations ($N = 10^3$ cells, $w = 0.5$, $m = 0.1$) with no therapy is plotted (black). The stochastic response to therapy is shown for 4 different time points in tumor progression. Therapy is designed to ``kill off" cancer cells with potency equal to their proportion. The first therapy (blue, beginning at $7\cdot10^3$ cell divisions) has the shortest time to total elimination of cancer cells. The second (red at $8.5\cdot 10^3$ cell) has a longer time to total elimination. The kill effect is diminished further for the third and fourth therapy (yellow, $10^4$ and purple, $1.15\cdot 10^4$ cell divisions) and some stochastic simulations are not able to eliminate all cancer cells but only delay saturation.}
\label{fig7}
\end{figure}

Figure \ref{fig7} shows the clear advantage of early stage therapy in our model system. We compare the effect of therapy given at an early stage, mid-stage, and late stages of the Gompertzian growth of the tumor. The black Gompertzian curve is the freely growing cancer cell population. The blue curve shows the cancer cell population diminishing to zero in time $\Delta t_1$ of continual therapy. The red curves shows the same therapy administered at a later stage in the growth phase (three stochastic simulations), diminishing to zero in time $\Delta t_2 > \Delta t_1$. The yellow curves show the therapy (five stochastic simulations) administered even later, diminishing in time $\Delta t_3 > \Delta t_2 > \Delta t_1$. In one of the simulations, the growing cancer cell population overcomes the killing effect of the therapy. The purple curves shown in the figure all are late stage therapies which do not kill the full population of cancer cells. In these cases, the growth of the cancer cell population outstrips the ability of the simulated therapy to kill them off.  Clearly, the earlier in the growth phase the therapy is administered, the less time is required to kill off the full population of cancer cells. But since the cancer cell population grows at very different rates throughout the full Gompertzian history of the developing tumor, this balance of kill cycles based on dose density and accelerated growth due to fitness advantage needs to be quantitatively determined.  

An established empirical law which relates drug dose density to its effectiveness in killing off cancer cells is known as the `log-kill' law \cite{bib51}. The log kill law states that a given dose of chemotherapy kills the same fraction of tumor cells (as opposed to the same number of tumor cells), regardless of the size of the tumor at the time the therapy is administered \cite{bib51}, a consequence of exponential growth with a constant growth rate.  This effect is best illustrated on a dose-response curve, plotting the dose density, $D$, with respect to the probability of tumor cell survival, $P_S$. The dose density is simply the product of the drug concentration, $c$, and the time over which the therapy is administered, $t$:

\begin{equation}
D = c \cdot t \label{eqn12}
\end{equation}

\noindent Thus, the log-kill law states the following:

\begin{equation}
\log (P_S) = - \beta D \label{eqn13}
\end{equation} 

\begin{figure}[ht!]
\begin{center}
\begin{subfigure}{.45\textwidth}
  \centering
  \noindent \includegraphics[width=1.0\linewidth]{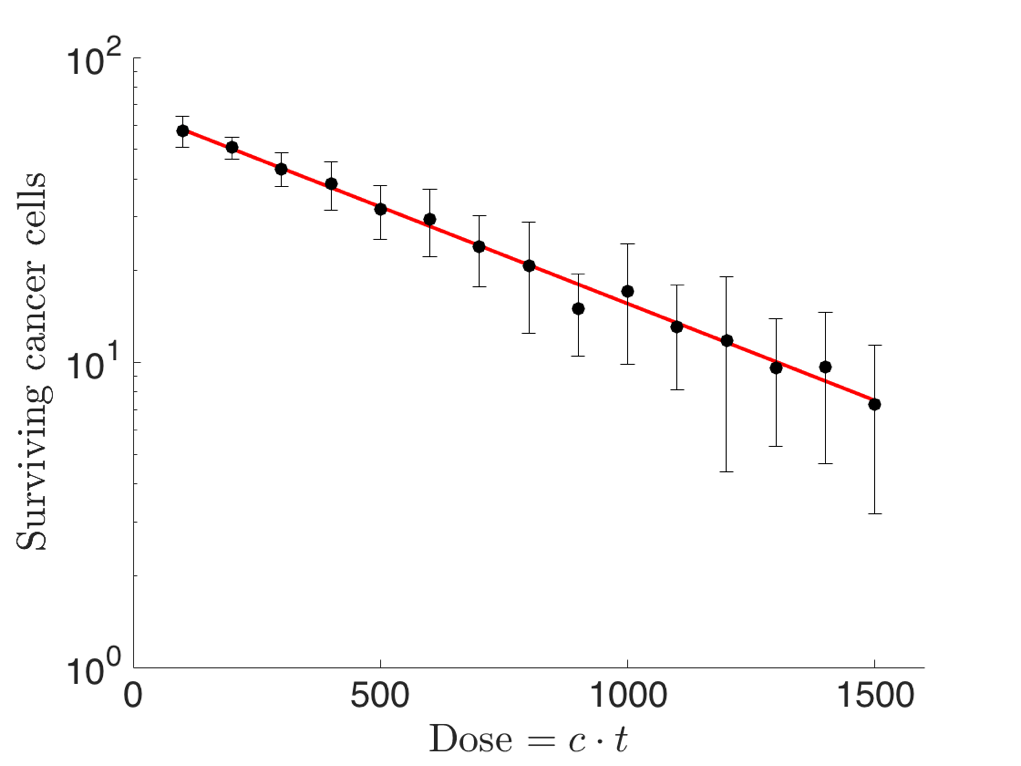}
  \caption{}{}
  \label{fig8:a}
\end{subfigure}
\begin{subfigure}{.45\textwidth}
  \centering
  \noindent \includegraphics[width=1.0\linewidth]{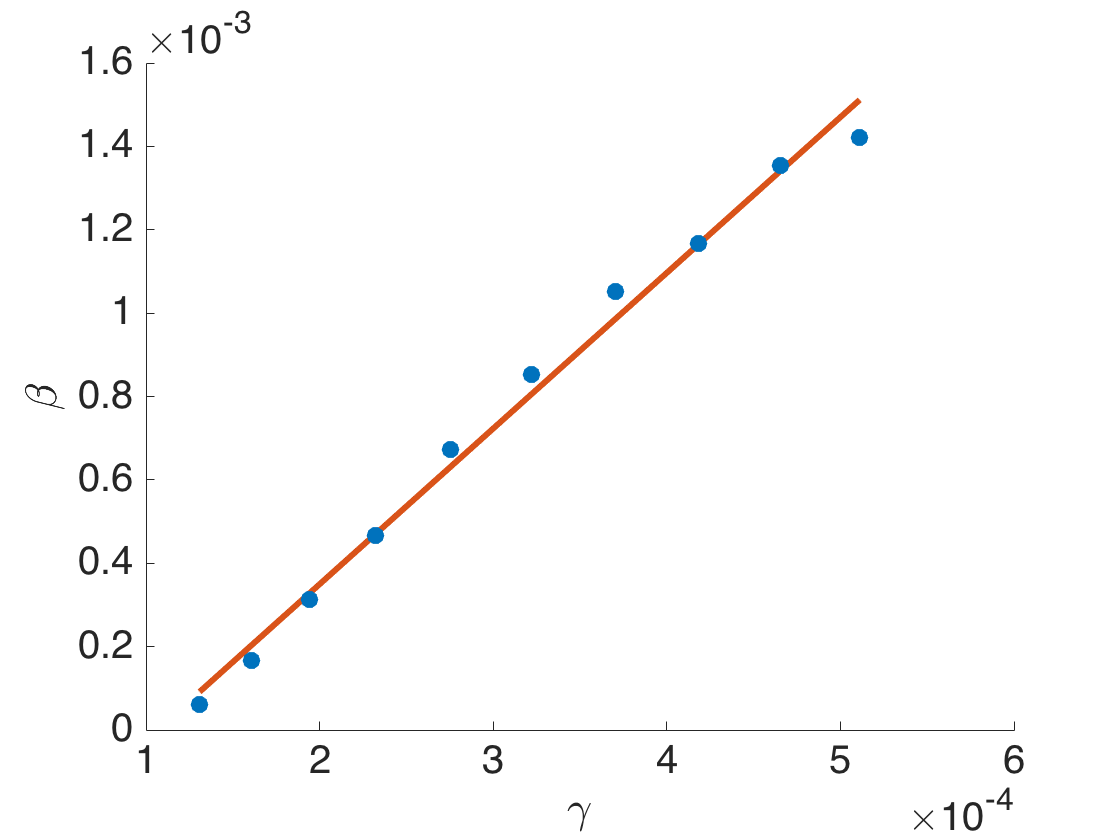}
  \caption{}{}
  \label{fig8:b}
\end{subfigure}
\end{center}
\caption{Growth-dependent tumor regression --- (a) Average of 25 stochastic simulations ($N = 10^3$ cells, $w = 0.5$, $m = 0.1$) where therapy is administered with varied dose densities ($0 \leq D \leq 1500$, $t = 100$). The rate of tumor regression, $\beta$, the slope of the red curve fit, is constant for a short period of simulated time where the tumor growth rate is approximately constant. This is the log-kill law.  (b) The process to find the rate of tumor regression, $\beta$, is repeated for a range of instantaneous tumor growth rates ($1\cdot 10^{-4} \leq \gamma \leq 6\cdot 10^{-4}$).  The Norton-Simon hypothesis predicts that $\beta$ is proportional to $\gamma$, indicated by a linear fit (red).}
\label{fig8:fig8}
\end{figure}

As an example, if there are 1000 cancer cells in a population, and the first therapy dose kills off 90\% of them, then after the first round of therapy there will be 100 cancer cells remaining. If a second round of therapy is administered, exactly as the first round, starting soon enough so that no new cancer cells have formed, then this next round will also kill off 90\% of the cells, leaving 10 cells, and so on for each future round of therapy. In a sense, since the first round killed 900 cells, while the second identical round killed only 90 cells, the population gets increasingly more difficult to kill off using the same treatment on each cycle. The log-kill law, a fundamentally static law (doesn't say anything about the relationship of the fraction of cells killed vs. the growth rate of the tumor), is verified in our model system, as shown in Figure \ref{fig8:a}. On the x-axis, we increase the dose density D, and we plot the number of surviving cancer cells. The slope of this straight line (verifying the log-kill law) can be thought of as the rate of regression of the tumor, $\beta$.

So how is the rate of regression, $\beta$, related to the growth rate of the tumor, $\gamma$? This is relevant, since we know from the shape of the Gompertzian curve, the growth rate is highest (exponential) at the beginning stage of tumor development and lowest at the late saturation stage. The Norton-Simon hypothesis \cite{bib41,bib42,bib43} states that the rate of regression is proportional to the instantaneous growth rate for an untreated tumor of that size at the time therapy is first administered. Faster growing tumors (early stage) should show higher rates of regression than more slowly growing tumors (late stage). This hypothesis is also verified in our model system, and shown clearly in Figure \ref{fig8:b}. The reality of this growth-dependent tumor regression rate effect (where early stage faster growing tumors are more vulnerable to therapy than later stage more slowly growing tumors) combined with the fact that the first round of therapy is more effective than future rounds of identical therapies, dramatically reinforces the need to administer drug treatment early in tumor progression when growth rates are high and there are fewer cancer cells to kill off. As drug concentration, $c$, is kept constant, the effect of treatment is reduced at later times in tumor development (Figure \ref{fig7}) until the same drug concentration is simply unable to overcome the tumor growth rate.

\section{Markov dissemination and progression patterns}

So how do these molecular and cellular growth details manifest themselves on the larger scales associated with metastatic progression patterns in patients? Despite the fact that disease progression patterns can vary from patient to patient, if a sufficiently large cohort of patients with similar characteristics is tracked over the course of the disease, statistical patterns emerge and can be exploited to build dynamical models of large scale progression. This lies at the heart of the models described in \cite{bib37,bib38,bib39} for lung cancer progression, and \cite{bib39,bib40} for breast cancer progression. 

\begin{figure}[!ht]
\begin{center}
\includegraphics[width=0.9\textwidth]{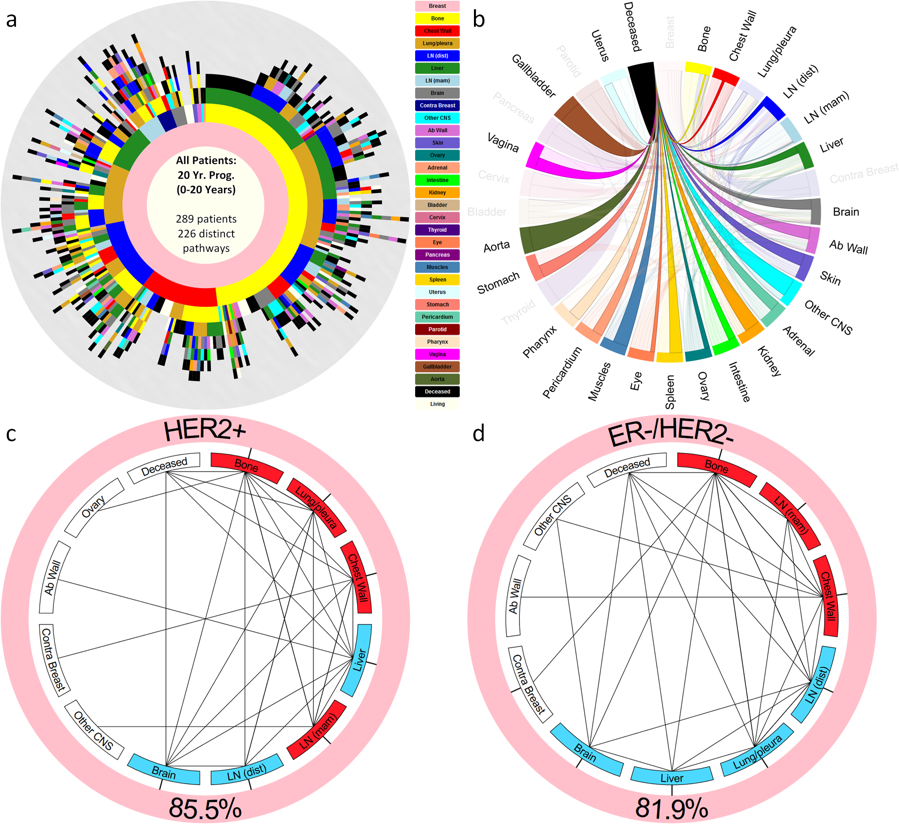}
\end{center}
\caption{Spatiotemporal patterns of breast cancer metastasis --- (a) Tree-ring diagram depicting all the paths in the clinical cohort over a 20-year period. (b) Markov chain network depicting transition probabilities from patients last metastatic tumor to deceased. (c) Reduced Markov chain diagram for sub-population of Her2+ patients. Red sites are spreader sites, blue sites are sponge sites. Note that bone is the main spreader. (d) Reduced Markov chain diagram for sub-population of ER-/Her2- patients. Red sites are spreader sites, blue sites are sponge sites. Note that bone is the main spreader, but Lung/pleura switches from being a spreader for Her2+ patients, to being a sponge for ER-/Her2- patient.}
\label{fig9}
\end{figure}

As an example of the kinds of whole-body scale models that can be built, consider first the tree-ring diagram shown in Figure \ref{fig9}a. The diagram encapsulates the entire progression history of a cohort of 289 primary breast cancer patients tracked at the Memorial Sloan Kettering Cancer Center for a 20 year period. All of the patients entered the cohort with a primary breast tumor, but no metastatic tumors. The inner ring, shown in pink, represents this cohort when they entered the study. As time progresses, the rings grow out, surrounding the inner breast ring. The first ring out shows the metastatic tumor distribution associated with first recurrence. The sector sizes represent the percentage of patients in this group. Likewise, the second ring out represents the distribution of tumors on second recurrence, and so forth for the further rings out. Hence, subsequent rings outward represent the tumor distributions as time progresses, with each patient history depicted on a ray going out from the center of the ring diagram. We caution that despite our usage of the term `tree-ring' diagrams for these representations, the thickness of the rings are all equal, hence do not reflect the time between subsequent recurrences (timescales of progression are documented and modeled in \cite{bib40}).  The power of the diagrams is that in one quick glance, one gains an appreciation for the statistical complexity of the disease \cite{bib39,bib40}. From them, one can also calculate the probability of the disease `transitioning' from one site to another as the disease progresses (called transition probabilities). These can then be used to create a single Markov transition matrix for each cancer type \cite{bib39}, which quantitatively encodes much of the information associated with the disease. Figure 9b shows the Markov transition graph from the last metastatic site to the deceased state for the cohort from Figure \ref{fig9}a. The sites are ordered clockwise from the most probably last metastatic site, to the least probable.

Figures \ref{fig9}c and \ref{fig9}d show reduced Markov diagrams \cite{bib39, bib40} for two specific important sub-groups of breast cancer patients, Her2+ patients, and ER-/Her2- patients. Generally speaking, Her2+ patients have the poorest prognosis. The red sites in these reduced diagrams (bone, lung/pleura, chest wall, LN (mam)) in Figure \ref{fig9}c, and bone, LN (mam), chest wall in Figure 9d are the spreaders associated with these groups. The blue sites (liver, LN (dist), brain for Fig. \ref{fig9}c; LN (dist), lung/pleura, liver, brain for Fig. \ref{fig9}d) are the sponges \cite{bib37,bib38,bib39}. It is interesting to note that lung/pleura switches from a spreader in the Her2+ sub-group to a sponge in the ER-/Her2- sub-group, suggesting a possible biological difference of the site in the different groups that correlates with different survival probabilities. 

\section{Mathematical modeling and tumor analytics}

It is important to keep in mind that no mathematical model captures all aspects of reality, so choices must be made which involve prioritizing the features that are most essential in capturing the essence of a complex process and which are not. Most experts now agree that the evolutionary processes in a tumor played out among subpopulations of competing cells are key to understanding aspects of growth and resistance to chemotherapy, which will ultimately lead the way toward a quantitative understanding of tumor growth and cancer progression \cite{bib31,bib59,bib60}. The paradigm of the cancer cell subpopulation and the healthy cell subpopulation competing as the defectors and cooperators in a Prisoner's Dilemma evolutionary game has been useful in obtaining a quantitative handle on many of these processes and frames the problem in an intuitive yet predictive way.  

Nonetheless, the mathematical `taste' of the modeler plays a role in what techniques are selected and ultimately where the spotlight shines. This fact makes clinicians uncomfortable and can lead to deep suspicion of the mathematical modeling enterprise as a whole.  Aren't the outcomes and predictions of mathematical models a straightforward consequence of the model assumptions? Once those choices are made, isn't the cake already baked? So why should we be surprised if you tell us it tastes good? Why not simply use tried and true statistical tools like regression methods to curve-fit the data directly, with no built in assumptions,  and be satisfied with uncovering correlations and trends? Clinicians (and experimentalists, in general) feel that they are dealing directly with reality, so why mess around with `toy' systems based on possibly `ad hoc' or incorrect assumptions that create artificial realities that may or may not be relevant? To a theoretician, calling their assumptions ad hoc, as opposed to natural, is as insulting as calling a clinician sloppy and uncaring (try this for yourself at the next conference you go to! But please use the term `somewhat ad hoc' to lessen the blow.) And if you want to deliver an even harsher insult, you could comment that the model seems like an exercise in curve fitting. 

But the usefulness of mathematical models built on simplified assumptions is well established in the history of the physical sciences, as detailed beautifully in Peter Dear's book, {\em The Intelligibility of Nature: How Science Makes Sense of the World} \cite{bib9}. Bohr's simple model of the structure of the atom was crucial in moving the community forward towards a deeper understanding of cause and effect, and ultimately pushing others to develop more realistic atomic models. The same could be said for many other important, but ultimately discarded models of reality (e.g. the notion of aether used as a vehicle to understand the mysterious notion of action-at-a-distance \cite{bib9}) now relegated to footnotes in the history of the physical sciences. 

Lessons from this history highlight the importance of using the principle of Occam's razor (law of parsimony) as a heuristic guide in developing models: (1) keep things simple, but not too simple;  (2) see what can be explained by using a given set of assumptions, and try to identify what is either wrong or cannot be explained; (3) add complexity to the model, but do this carefully. Since ultimately, the model will always be wrong (with respect to some well chosen and specific new question being posed about a system), it is important that it be {\em useful as a vehicle of intelligibility} \cite{bib9} associated with the set of questions surrounding the phenomena it was built to explain. Answers to some new questions will be found using the model as a temporary crutch, and new questions will emerge in the process that had not yet been asked, as their relevance had never previously been realized. A new quantitative language will emerge in which aspects of the model will be associated with the underlying reality it is attempting to describe, predictions will be easier to frame and test, and shortcomings will be exposed.  In his famous article \cite{bib62}, Eugene Wigner writes compellingly that `the miracle of the appropriateness of the language of mathematics for the formulation of the laws of physics is a wonderful gift which we neither understand nor deserve. We should be grateful for it and hope that it will remain valid in future research and that it will extend, for better or for worse, to our pleasure, even though perhaps also to our bafflement, to wide branches of learning.'

In general, the more complex the model (as measured, for example, by the number of independent parameters associated with it), the less useful it will be, and the less likely it is to be adopted by the community at large. After all, if the model is as complex as the phenomena it was built to understand, why not stick with reality? Effective models can be thought of as {\em low-dimensional approximations of reality}, surrogates that help us bootstrap our way forward. They arise as the outcome of a complex balancing act between simplicity of the ingredients, and complexity of the reality the model is meant to describe. They generally do not arise in a vacuum, but are built in the context of informed and sustained discussions among people with different expertise. In the context of medical oncology, this means physical scientists developing ongoing interactions with clinical oncologists, radiologists, pathologists, molecular and cell biologists and other relevant medical specialists.

Appropriate data is a necessary ingredient in developing and testing any successful model, and treasure troves of medical data sit unexamined in patient files and government databases across the country waiting to be put to good use. There is no doubt that they are telling an interesting and important story that we have yet to fully understand. It is not currently possible for the computer to simulate all of the complex, relevant, and systemic ingredients at play to faithfully recreate all aspects of cancer progression and treatment response in patients. It is hard to imagine that a deep and actionable understanding can ever be obtained without the combined use of data, models, and computer simulations to help guide us and highlight some of the underlying causal mechanisms of this complex and deadly disease.


\section*{References}
\begin{thebibliography}{10}

\bibitem{bib1} {CSO Attalini, F Michor},
{\em Evolutionary theory of cancer}, Annals of NY Academy of Sci., 1168 (1) 23-51

\bibitem{bib2} {R Axelrod},
{\bf The Evolution of Cooperation}, 1984, Basic Books, New York.

\bibitem{bib3} {R Axelrod, DE Axelrod, K Pienta},
{\em Evolution of cooperation among tumor cells}, Proc. Natl. Acad. Sci. 103 (36), 13474-13479.

\bibitem{bib4} {D Basanta, A Deutsch},
{\em A game theoretic perspective on the somatic evolution of cancer}, 2008 Selected Topics in Cancer Modeling 1-16.

\bibitem{bib5} {R.A. Burrell, N. McGranahan, J. Bartek, C. Swanton},
{\em The causes and consequences of genetic heterogeneity in cancer evolution}, Nature Review Vol. 509 19 September 2013 338-345. Nature Insight: Tumour Heterogeneity 19 Sept 2013 Vol 501 No 7467

\bibitem{bib6} {T.M. Cover, J.A. Thomas},
{\bf Elements of Information Theory}, 2nd Ed. Wiley-Interscience, 2006.

\bibitem{bib7} {B Crespi, K Summers},
{\em Evolutionary biology of cancer}, Trends in Ecol. and Evol. 2005, Vol 20(10), 545-553.

\bibitem{bib8} {R. Dawkins},
{\bf The Selfish Gene}, Oxford University Press 1976.

\bibitem{bib9} {P. Dear},
{\bf The Intelligibility of Nature: How Science Makes Sense of the World}, U. Chicago Press, 2006.

\bibitem{bib10} {M. Doebeli, C. Hauert},
{\em Models of cooperation based on Prisoner's Dilemma and the Snowdrift game}, Ecology Lett. 8(7), 748-766 2005.

\bibitem{bib11} {M. Doebeli, C. Hauert, J. Killingback},
{\em The evolutionary origin of cooperators and defectors}, Science 306 (5697), 859-862, 2004.

\bibitem{bib12} {SA Frank},
{\bf Dynamics of Cancer: Incidence, Inheritance and Evolution}, 2007, Princeton University Press.

\bibitem{bib13} {S.A. Frank, M. Nowak},
{\em Problems of somatic mutation and cancer}, Bio. Essays 26,3 291-299 2004.

\bibitem{bib14} {RA Gatenby, TL Vincent},
{\em An evolutionary model of carcinogenesis}, Cancer Res. 2003 Oct 63(19): 6212-6220.

\bibitem{bib15} {RA Gatenby, TL Vincent},
{\em Application of quantitative models from population biology and evolutionary game theory to tumor therapeutic strategies}, Mol Cancer Ther 2003 Sept 2(9): 919-927.

\bibitem{bib16} {JH Goldie, AJ Coldman},
{\bf Drug Resistance in Cancer: Mechanisms and Models}, Cambridge University Press 1998.

\bibitem{bib17} {C. Hauert},11-44, 2008.
{\em Evolutionary dynamics, evolution from cellular to social scales}, 

\bibitem{bib18} {C. Hauert, G. Szabo},
{\em Game theory and physics}, Am. J. Phys. 73(5) 405-414, 2005.

\bibitem{bib19} {J Hofbauer, K Sigmund},
{\bf Evolutionary Games and Population Dynamics}, Cambridge University Press 1998.

\bibitem{bib20} {S Hummert, K Bohl, D Basanta  et al.},
{\em Evolutionary game theory: Cells as players}, Mol. BioSyst. 2014, 10, 3044-3065. 

\bibitem{bib21} {S Hummert, K Bohl, D Basanta et al.},
{\em Evolutionary game theory: Molecules as players}, Mol. BioSyst. 2014, 10, 3066-3074.

\bibitem{bib22} {Y Iwasa,  MA Nowak, F Michor},
{\em Evolution of resistance during clonal expansion}, Genetics April 2006, Vol 172, No. 4 2557- 2566.

\bibitem{bib23} {I. Kareva},
{\em Prisoner's dilemma in cancer metabolism}, PLoS ONE Dec. 2011 Vol. 6(12) e28576.

\bibitem{bib24} {W.S. Kendal},
{\em Gompertzian growth as a consequence of tumor heterogeneity}, Math. Biosciences 73: 103-107 1985.

\bibitem{bib25} {A.K. Laird},
{\em Dynamics of tumor growth}, Br J. Cancer, 1964 Sept 18(3) 490-502.

\bibitem{bib26} {A.K. Laird},
{\em  Dynamics of tumor growth: Comparison of growth rates and extrapolation of growth curve to one cell}, British J. Cancer, 1965

\bibitem{bib27} {Lieberman, E., Hauert, Ch., Nowak, M. A.},
{\em (2005) Evolutionary dynamics on graphs}, Nature 433 312-316.

\bibitem{bib28} {Martincorena I., Campbell PJ},
{\em (2015) Somatic mutations in cancer and normal cells}, Science 25 Sept. Vol. 349 Issue 6255 1483-1489.

\bibitem{bib29} {J. Maynard Smith},
{\em Evolutionary game theory}, Physica 22D 43-49 1986.

\bibitem{bib30} {J. Maynard Smith},
{\bf Evolution and The Theory of Games}, Cambridge University Press 1982.

\bibitem{bib31} {LM Merlo, JW Pepper, BJ Reid, CC Maley},
{\em ancer as an evolutionary and ecological process}, Nat Rev Cancer 2006 Dec 6(12): 924-35.

\bibitem{bib32} {F Michor},
{\em Evolutionary Dynamics of Cancer}, h.D. Thesis, Organismic and Evolutionary Biology, Harvard University, 2005.


\bibitem{bib33} {F. Michor, Y. Iwasa, M. Nowak},
{\em Dynamics of cancer progression}, Nature Rev. Cancer Vol. 4, March 2004, 197-205.

\bibitem{bib34} {J.F. Nash Jr.},
{\em Equilibrium points in N-person games}, Proc. Natl Acad Sci. USA 1950 Jan. 36(1), 48-49.

\bibitem{bib35} {J.F. Nash Jr.},
{\em Non-cooperative games}, The Annals of Math. Vol. 54, Issue 2 Sept. 1951 286-295


\bibitem{bib36} {}
{\em Nature Milestones Cancer}, April 2006.

\bibitem{bib37} {P.K. Newton, J. Mason, K. Bethel, L. Bazhenova, J. Nieva, P. Kuhn},
{\em A stochastic Markov chain model to describe lung cancer growth and metastasis}, PLoS one, Vol 7, 4, e34637 April 2012.


\bibitem{bib38} {P.K. Newton, J. Mason, K. Bethel, L. Bazhenova, J. Nieva, L. Norton, P. Kuhn},
{\em Spreaders and sponges define metastasis in lung cancer: A Markov chain Monte Carlo mathematical model}, Cancer Res. 73(9) May 1 2013.



\bibitem{bib39} {	P.K. Newton, J. Mason, B. Hurt, K. Bethel, L. Bazhenova, J. Nieva, P. Kuhn},
{\em Entropy, complexity, and Markov diagrams for random walk cancer models}, Nature Sci. Rep. 4:7558 2014.



\bibitem{bib40} {P.K. Newton, J. Mason, N. Venkatappa, M.S. Jochelson, B. Hurt, J. Nieva, E. Comen, L. Norton, P. Kuhn},
{\em Spatiotemporal progression of metastatic breast cancer: A Markov chain model highlighting the role of early metastatic sites}, npj Breast Cancer 1, 15018, 21 Oct. 2015.


\bibitem{bib41} {Norton L and Simon R},
{\em (1976) Tumor size, sensitivity to therapy and the design of treatment protocols}, Cancer Treat Rep 61: 1307–1317


\bibitem{bib42} {Norton L and Simon R},
{\em (1977) The growth curve of an experimental solid tumor following radiotherapy}, J Natl Cancer Inst 58: 1735–1741

\bibitem{bib43} {Norton L and Simon R},
{\em (1986) The Norton-Simon hypothesis revisited}, Cancer Treat Rep 70: 163–169 


\bibitem{bib44} {MA Nowak},
{\bf Evolutionary Dynamics: Exploring the Equations of Life}, 2006, Harvard University Press.


\bibitem{bib45} {M. Nowak},
{\em Stochastic strategies in the Prisoner's dilemma}, Theor. Pop. Bio. Vol. 38, Issue 1 93-112, Aug. 1990.


\bibitem{bib46} {M. Nowak, R.M. May},
{\em Evolutionary games and spatial chaos}, Nature 359 (6398) 826-829, 1992.


\bibitem{bib47} {M. Nowak, K. Sigmund},
{\em Evolutionary dynamics of biological games}, Science, 303 (5659), 793-799 2004.


\bibitem{bib48} {PC Nowell},
{\em The clonal evolution of tumor cell populations}, Science 1976 Vol. 194, No. 4260, 23-28.

\bibitem{bib49} {O Podlaha, M Riester, S De, F Michor},
{\em Evolution of the cancer genome}, Trends in Genetics, 28(4) 155-163.


\bibitem{bib50} {S Schuster, JU Kreft, A Schroeter, T Pfeiffer},
{\em Use of game theoretical methods in biochemistry and biophysics}, J. Biol. Phys. 2008 34:1-17.


\bibitem{bib51} {Skipper HE},
{\em (1986) Laboratory models: some historical perspectives}, Cancer Treat Rep 70: 3–7.

\bibitem{bib52} {J.S. Spratt, J.A. Spratt},
{\em What is breast cancer doing before we can detect it?}, J. Surg. Onc. 1985 Nov 30 (3) 156-160.


\bibitem{bib53} {C. Swanton},
{\em Intratumor heterogeneity: evolution through space and time}, Cancer Res. 72(19) 4875-4882 Oct. 1 2012.



\bibitem{bib54} {IPM Tomlinson},
{\em Game theory models of interactions between tumour cells}, Eur. J. Cancer, 2003 Vol. 33, N9 1495-1500


\bibitem{bib55} {A. Traulsen, C. Hauert},
{\em Stochastic evolutionary game dynamics}, Reviews of nonlinear dynamics and complexity, 2, 25-61, 2009.

\bibitem{bib56} {T. Vincent},
{\em Carcinogenesis as an evolutionary game}, Adv. In Complex Sys. 2006 9(4) 369-382.


\bibitem{bib57} {J. von Neumann, O. Morgenstern},
{\bf Theory of Games and Economic Behavior}, Princeton University Press, 1944.




\bibitem{bib58} {J. W. Weibull},
{\bf Evolutionary Game Theory}, MIT Press, 1997.


\bibitem{bib59} {R.A. Weinberg},
{\bf One Renegade Cell: How Cancer Begins}, Basic Books 1998.


\bibitem{bib60} {R.A. Weinberg},
{\bf The Biology of Cancer}, Garland Science, 2007.

\bibitem{bib61} {J. West, Z. Hasnain, P. Macklin, J. Mason, P.K. Newton},
{\em An evolutionary model of tumor cell kinetics and the emergence of molecular heterogeneity and Gompertzian growth}, NOTE = {preprint, \url{http://arxiv.org/abs/1512.04590}}




\bibitem{bib62} {E. Wigner},
{\em The unreasonable effectiveness of mathematics in the natural sciences}, Comm. Pure and Appl. Math. Vol. 13, No. 1 Feb. 1960.


















\end{thebibliography}
\end{document}